\documentclass[prd,onecolumn]{revtex4}

\usepackage{amsmath,amssymb}
\usepackage{multirow}
\usepackage{bbold}
\usepackage{graphicx}
\usepackage{subfigure}
\newcommand{\be}{\begin{equation}}
\newcommand{\ee}{\end{equation}}

\newcommand{\lp}{\left(}
\newcommand{\rp}{\right)}
\newcommand{\bb}{\begin{bmatrix}}
\newcommand{\eb}{\end{bmatrix}}
\DeclareMathOperator{\Tr}{Tr}

\begin{document}


\title{Results of the First IPTA Closed Mock Data Challenge}

\author{J. Ellis}
\affiliation{Department of Physics, University of Wisconsin-Milwaukee,
 Milwaukee, WI 53201, USA}

\author{X. Siemens}
\affiliation{Department of Physics, University of Wisconsin-Milwaukee,
Milwaukee, WI 53201, USA}

\author{S. Chamberlin}
\affiliation{Department of Physics, University of Wisconsin-Milwaukee,
Milwaukee, WI 53201, USA}

\begin{abstract}
The 2012 International Pulsar Timing Array (IPTA) Mock Data Challenge (MDC) is designed to test current Gravitational Wave (GW) detection algorithms. Here we will briefly outline two detection algorithms for a stochastic background of gravitational waves, namely, a first-order likelihood method and an optimal statistic method and present our results from the closed MDC data sets.  
\end{abstract}

\maketitle

\section{Generating the Residuals}
The first step in any data analysis pipeline used in this paper is to generate the residuals and make sure that all of the fits converged. Here we use \textsc{tempo2} with the \texttt{general2} plugin to generate the residuals. For our analysis we will also need the design matrix for each pulsar which is obtained using the \texttt{designmatrix} plugin developed by R. van Haasteren and modified by J. Ellis. From inspection of the residuals, it is immediately obvious that all three datasets contain strong red noise in many of the residuals. We can also see ``by-eye'' that the residuals of dataset 1 have similar errorbars and that the errorbars of datasets 2 and 3 seem to vary quite a bit from pulsar to pulsar.

\section{Noise Estimation}

The next step in our analysis pipelines is to estimate the red and white noise levels in each set of residuals. A detailed description of the process is in preparation \citep{esd+12}. Here we will simply review the method and present the results from the Mock Data Challenge. We model the noise in the data as a sum of three processes: a red component with a power law power spectrum, a systematic white noise component that multiplies the residual error bars (EFAC), and an extra white noise component independent of the error bars (EQUAD). We work in the time domain, so these noise processes can be completely described by their covariance matrices. Here we model the \emph{pre-fit} covariance matrix as
\be
\Sigma_{y}=\langle \mathbf{y}\mathbf{y}^{T} \rangle=C_{y}^{\rm red}+C_{y}^{\rm EFAC}+C_{y}^{\rm EQUAD}.
\ee
Details of the various components and transformations into a \emph{post-fit} basis will be described elsewhere \citep{esd+12}. For this discussion it is only important to note that we maximize the likelihood function
\be
\mathcal{L}(\vec\theta|\mathbf{r})=\frac{1}{\sqrt{\det(2 \pi \Sigma_r)}}\exp(-\frac{1}{2}\mathbf{r}^T\Sigma_r^{-1}\mathbf{r})
\ee
over the parameters $\vec\theta=\{\mathcal{A},\gamma,\mathcal{F},\mathcal{Q}\}$, where $\mathcal{A}$ and $\gamma$ are the amplitude and spectral index of the red noise power spectrum and $\mathcal{F}$ and $\mathcal{Q}$ are the EFAC and EQUAD parameters, respectively. In the above expression, $\Sigma_{r}$ and $\mathbf{r}$ are the \emph{post-fit} covariance matrix and residuals, respectively.

\begin{figure*}[!h]
  \begin{center}
	\includegraphics[scale=1.0]{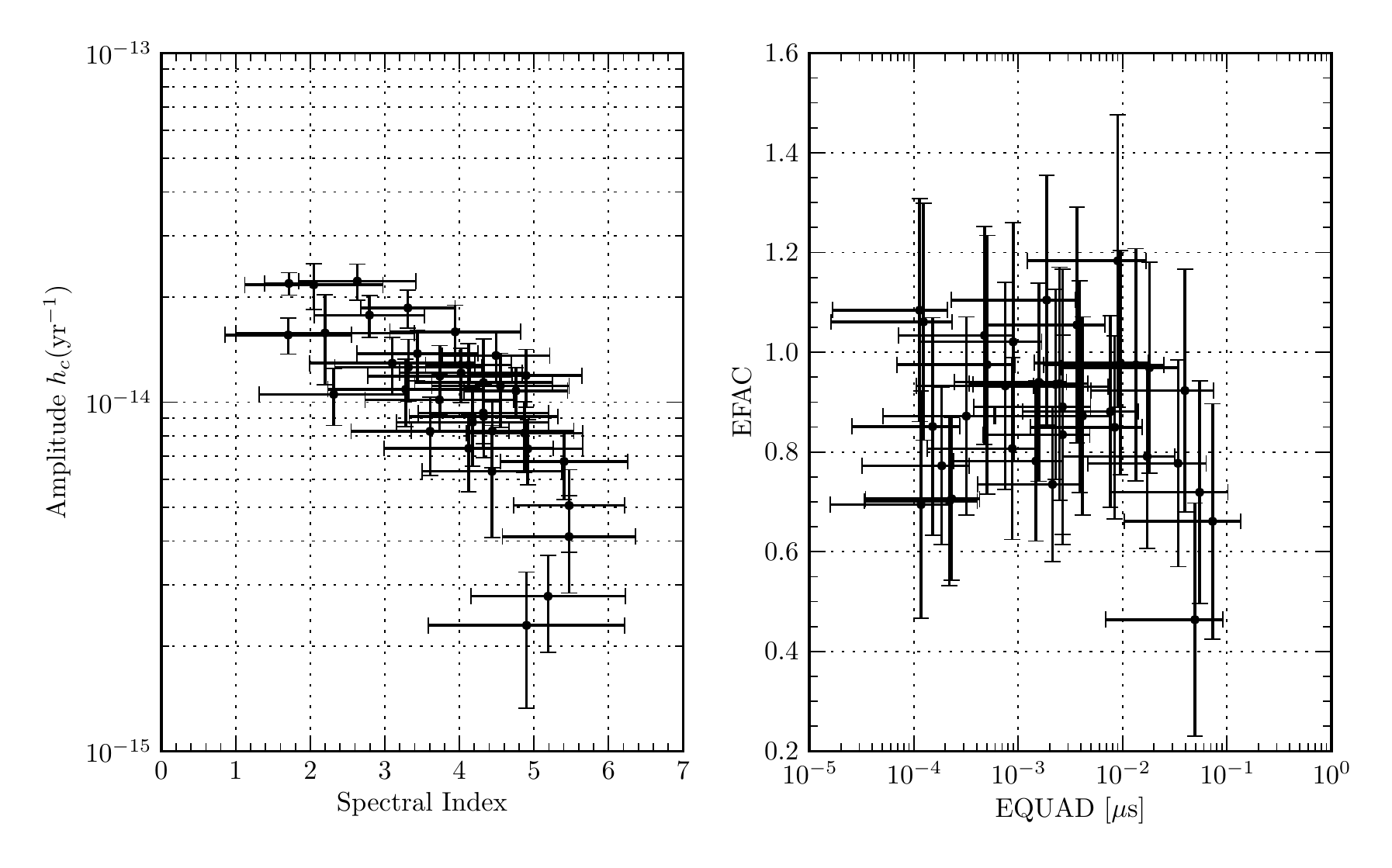}
  \end{center}
  \caption{Results of our noise estimation algorithm on individual dataset 1 pulsars. The left plot is the amplitude of a red noise signal converted to units of GWB strain vs. the power spectral index. The plot on the right is the EFAC parameter plotted against the EQUAD parameter in microseconds.}
\label{fig:noiseMDC1}
\end{figure*}

\begin{figure*}[!h]
  \begin{center}
	\includegraphics[scale=1.0]{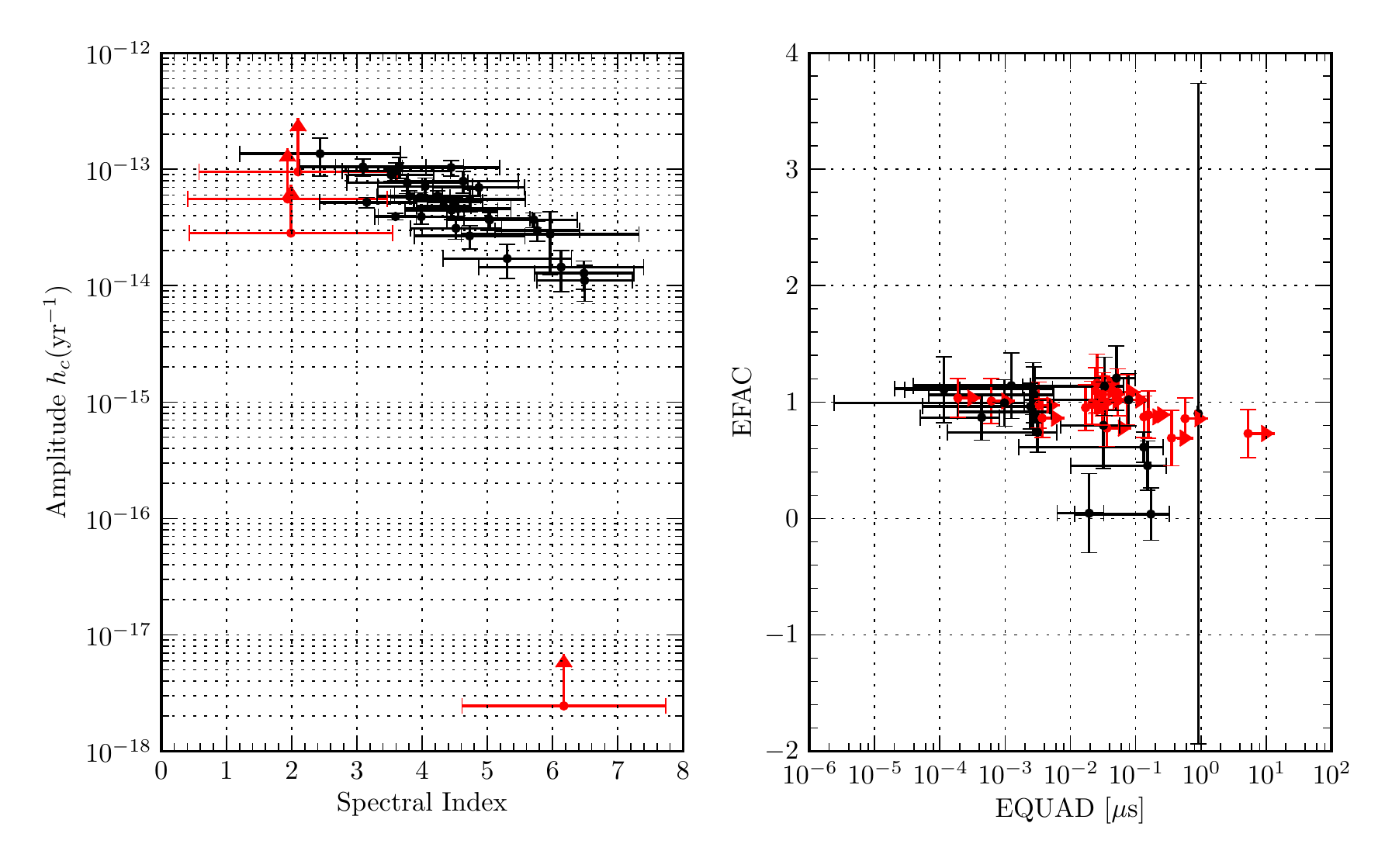}
  \end{center}
  \caption{Results of our noise estimation algorithm on individual dataset 2 pulsars. The left plot is the amplitude of a red noise signal converted to units of GWB strain vs. the power spectral index. Note that the red points show upper limits on the amplitude and EQUAD parameters respectively. This is done because these points are consistent with 0 at the 1-sigma level. Here, 32 of the 36 pulsars show evidence for red noise in the individual case indicating the presence of a strong red noise signal in nearly all pulsars. The plot on the right is the EFAC parameter plotted against the EQUAD parameter in microseconds.}
\label{fig:noiseMDC2}
\end{figure*}

\begin{figure*}[!h]
  \begin{center}
	\includegraphics[scale=1.0]{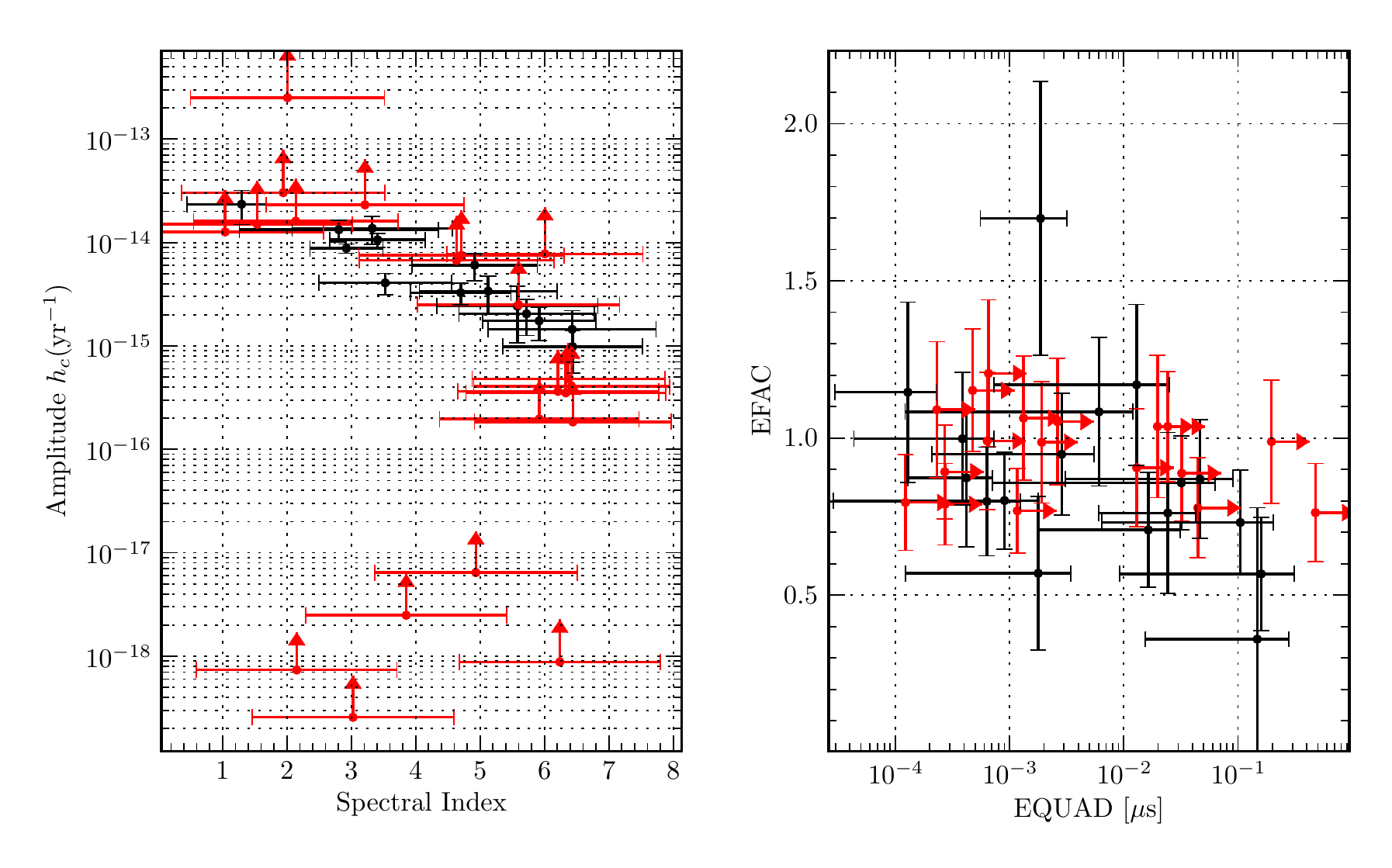}
  \end{center}
  \caption{Results of our noise estimation algorithm on individual dataset 3 pulsars. The left plot is the amplitude of a red noise signal converted to units of GWB strain vs. the power spectral index.  Note that the red points show upper limits on the amplitude and EQUAD parameters respectively.  This is done because these points are consistent with 0 at the 1-sigma level. Here, only 15 of the 36 pulsars show evidence for red noise in the individual case indicating that many pulsars are white noise dominated. The plot on the right is the EFAC parameter plotted against the EQUAD parameter in microseconds.}
\label{fig:noiseMDC3}
\end{figure*}

We have analyzed each pulsar from each dataset and compiled the results in Figures \ref{fig:noiseMDC1}--\ref{fig:noiseMDC3}. For this four parameter search, typical runtimes are on the order of 30 minutes per pulsar. When run in parallel using MPI, these runtimes decrease to a few minutes per pulsar.
In these figures we plot the red noise amplitude in familiar GWB strain units vs. the power spectral index. For example, in these units a GWB from supermassive black hole binaries SMBHBs with a characteristic spectral index of -2/3 will have a power spectral index of 13/3. We also plot the EFAC parameter vs. the EQUAD parameter. In general if the white noise were completely described by the errorbars then the EFAC parameter is unity and the EQUAD parameter is 0. However, since these datasets have equal errorbars on all TOAs, EFAC and EQUAD are degenerate and may not result in the above situation. However, from these results we can conclude that there is no evidence for any white noise other than radiometer noise contained in the error bars. Red points indicate that the amplitude or EQUAD is consistent with 0 at the one sigma level and we simply plot upper limits on those parameters (red triangles). Here we notice that MDC 1 has strong red noise in all the residuals clustered around an Amplitude of $A\sim 1\times 10^{-14}$ and a spectral index of $\gamma\sim 4.0$. MDC 2 is centered around a higher amplitude and similar spectral index but with a larger spread. In contrast, many of the pulsars in MDC 3 show little evidence for red noise, with quite a large spread in both spectral index and amplitude.

\section{Analysis Methods}

\subsection{Optimal Statistic}
The optimal statistic (OS) used for this work is based on that published in \citet{abc+09} and will be further discussed in a future work. Here we will briefly review the algorithm. In general, we wish to construct a statistic that gives some measure of the significance of the expected (Hellings-Downs curve) correlations among pulsars giving larger weight to those residuals with lower noise levels. It is also useful to construct this statistic, such that, on average its value is the amplitude of the GWB for some given spectral index. It is shown in \cite{abc+09} that this statistic is both an optimal filter and maximizes the likelihood function in the low-signal limit. The optimal statistic can be written as
\be
\hat{\Omega}=\frac{\sum_{I,J}r_{I}^{T}P_{I}^{-1}S_{IJ}P_{J}^{-1}r_{J}}{\sum_{I,J}{\rm tr}\left[P_{I}^{-1}S_{IJ}P_{J}^{-1}S_{JI}\right]}
\ee
with an SNR of 
\be
\hat{\rho}=\frac{\sum_{I,J}r_{I}^{T}P_{I}^{-1}S_{IJ}P_{J}^{-1}r_{J}}{\sqrt{\sum_{I,J}{\rm tr}\left[P_{I}^{-1}S_{IJ}P_{J}^{-1}S_{JI}\right]}},
\ee
where $\sum_{I,J}$ denotes the sum over pulsar pairs, $r_{I}$, $P_{I}$, and $S_{IJ}$ are the residuals for pulsar $I$, the auto-covariance matrix of pulsar $I$ and the amplitude free cross covariance matrix of pulsar pair $IJ$. We use the amplitude free cross covariance matrix to ensure that $\langle \hat{\Omega} \rangle=\Omega_{\rm true}$ under the assumption that $\langle r_{I}r_{J}^{T}\rangle=\Omega_{\rm true}S_{IJ}$. The SNR used here gives a measure of how strong the \emph{correlated} signal is compared to an \emph{uncorrelated} signal of the same strength.

For this analysis we use two versions of the optimal statistic: the zero-order and the iterative optimal statistic. The zero-order optimal statistic consists of estimating the maximum likelihood noise parameters, as discussed above, and forming the auto-covariance matrices as input into the code. In general, this method will give a biased result for the estimate of the GWB amplitude $\Omega$ because we are ignoring the presence of a common GWB signal in all pulsars. As we can see from Figures \ref{fig:noiseMDC1}--\ref{fig:noiseMDC3}, there is quite a large spread on the red noise parameters and this large spread adds to the bias in the optimal statistic. However, if we iterate this procedure we can obtain a much better estimate of $\Omega$ and a higher SNR. The iterative procedure is as follows: 
\begin{enumerate}
\item Run the zero-order likelihood with the auto-covariance matrices produced from the noise estimation procedure to obtain an estimate of the GWB amplitude, $\Omega_{0}$ (here we assume a known spectral index in the construction of the cross-covariance matrices).
\item Now fix the spectral index for both the auto and cross covariance matrices and use the estimate $\Omega_{0}$ from the previous step to produce new auto-covariance matrices.
\item Run the optimal statistic code again, now with the new auto-covariance matrices and produce a new estimate of the amplitude $\Omega_{i}$
\item Repeat these steps until the new estimate of the amplitude changes from the previous estimate by some amount $\varepsilon$\footnote{For this analysis, $\varepsilon$ was chosen to be 0.005.}.
\end{enumerate}
It is important to note that for the iterative method to be valid we must have a priori knowledge about any red noise intrinsic to the pulsars. However, for the MDC datasets, no pulsars show evidence of intrinsic red noise, so it is a good approximation to assume that the only red noise source is the GWB. Results for both the zero-order and iterative optimal statistics are shown in Table \ref{tab:results} and will be discussed in the final section.

\subsection{First-Order Likelihood}
The first-order likelihood combines elements of \citet{hlm+09} and \cite{abc+09} and will be published in a future paper. For a pulsar timing array with $M$ pulsars we define the probability distribution function of 
the presumed Gaussian noise as multivariate Gaussian\footnote{Here we will assume that all noise processes are stochastic and that there are no deterministic signals present in the data}
\be
p(\mathbf{r})=\frac{1}{\sqrt{\det2\pi\boldsymbol{\Sigma}}}\exp\lp -\frac{1}{2}\mathbf{r}^{T}
\boldsymbol{\Sigma}^{-1}\mathbf{r} \rp,
\ee
where
\be
\mathbf{r}=\bb {r}_{1} \\ {r}_{2}\\ \vdots \\ {r}_{M} \eb
\ee 
is a vector of the residual time-series, $r_{\alpha}(t)$, for all pulsars,
\be
\label{eq:cov}
\boldsymbol{\Sigma}=\bb  P_{1} & S_{12} & \hdots & S_{1M}\\ 
S_{21} & P_{2} & \hdots & S_{2M}\\
\vdots & \vdots & \ddots & \vdots\\
S_{M1} & S_{M2} & \hdots & P_{M}\eb
\ee
is a multivariate  block covariance matrix, and 
\begin{align}
P_{{\alpha}}&=\langle r_{\alpha}r_{\alpha}^{T}\rangle\\
S_{\alpha\beta}&=\langle r_{\alpha}r_{\beta}^{T}\rangle\big|_{\alpha\ne \beta},
\end{align}
are the auto-covariance and cross-covariance matrices, respectively, for each set of residuals. 

Here we are interested in measuring the spectral index, $\gamma_{\rm gw}$, and amplitude, $\Omega$,  of the stochastic background. These parameters will be common among all pulsars, however, as mentioned above, each pulsar will have intrinsic noise parameters as well: an amplitude $A_{\alpha}$ and spectral index $\gamma_{\alpha}$ for a power law red noise process, and EFAC and EQUAD parameters, $\mathcal{F}_{\alpha}$ and $\mathcal{Q}_{\alpha}$, for white noise processes. Therefore, we write our auto-covariance as a sum of a common GWB term and a pulsar dependent term
\be
P_{\alpha}=R_{\alpha}+S_{a\alpha},
\ee
where $R_{\alpha}$ is the intrinsic noise auto-covariance matrix and $S_{a\alpha}$ is the common GWB auto-covariance matrix. It is convenient to work in a block matrix notation where
\be
\mathbf{\Sigma}=\mathbf{P}+\mathbf{S}_{c},
\ee
where $\mathbf{P}$ is a block diagonal matrix with diagonals $P_{\alpha}$ and $\mathbf{S}_{c}$ is block matrix with off diagonals $S_{\alpha\beta}$ and zero matrices on the diagonal.

In general, we write the log likelihood function as
\be
\ln\,\mathcal{L}=-\frac{1}{2}\left[ \Tr\, \ln \mathbf{\Sigma} +\mathbf{r}^{T}\mathbf{\Sigma}^{-1}\mathbf{r} \right],
\ee
where we have used the fact that $\ln \det(A)=\Tr \ln(A)$. 
\begin{figure*}[!h]
  \begin{center}
	\includegraphics[scale=0.56]{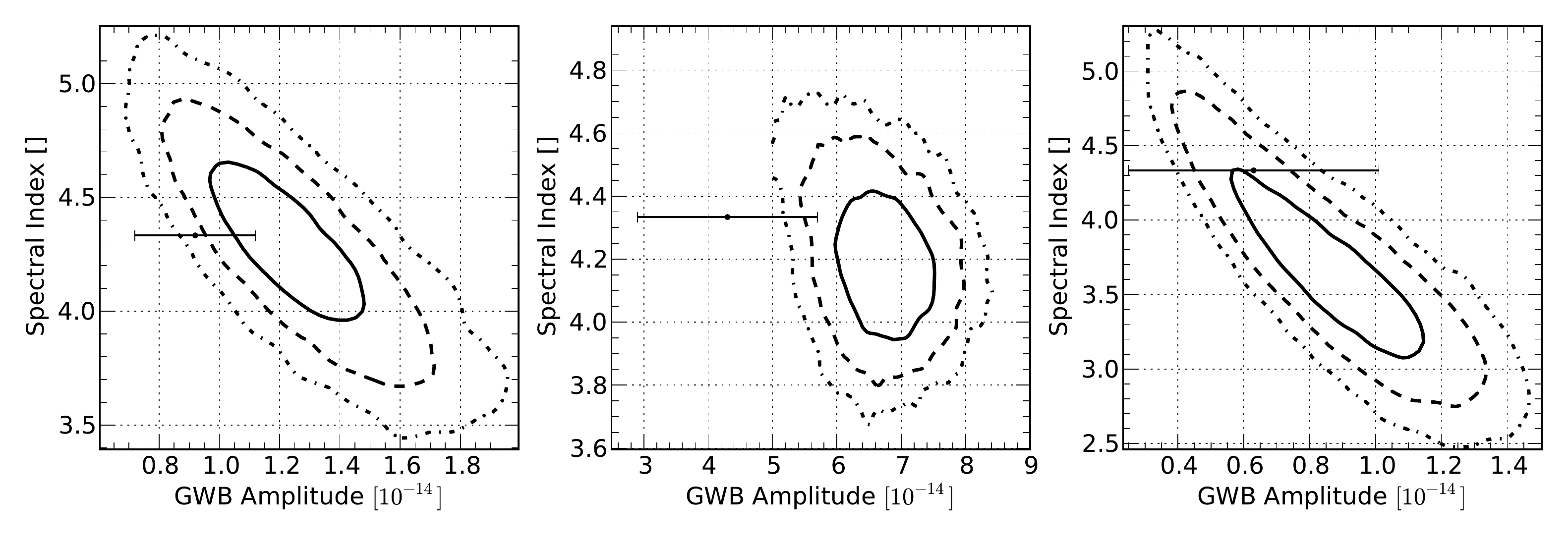}
  \end{center}
  \caption{(from left to right) Results from our first-order likelihood method run on the ``best 12'' pulsars for closed mock datasets 1, 2 and 3. Here the solid, dashed and dash-dot lines are the 68\%, 95\%, and 99\% contours, respectively.  The points with errorbars are the estimates of the amplitude obtained from the iterative optimal statistic with 1-sigma uncertainties.}
\label{fig:likeContours}
\end{figure*}
In practice the matrix $\mathbf{\Sigma}$ is quite large and therefore, computationally prohibitive to invert. Since many multi-frequency residual datasets now have on the order of $10^{3}$ points, for many modern PTAs the matrix $\mathbf{\Sigma}$ will be of order $10^{4}\times 10^{4}$. To avoid inverting the full covariance matrix we can expand the inverse out in a Neumann series to obtain
\be
\mathbf{\Sigma}^{-1}=\mathbf{P}^{-1}-\mathbf{P}^{-1}\mathbf{S}_{c}\mathbf{P}^{-1}+\mathcal{O}(\epsilon^{2}).
\ee
where $\epsilon$ is an order parameter. Physically, $\epsilon$ can represent the amplitude of the GWB, $\Omega$, since it is a \emph{small} constant multiplier of the elements of $\mathbf{S}_{c}$  Since the diagonal elements of $\mathbf{\Sigma}$ will always be less than the diagonal terms, we can keep only up to first order in $\epsilon$. The determinant can also be approximated (to first order in $\epsilon$) as
\be
\ln \det\mathbf{\Sigma}=\Tr\ln\mathbf{P}+\mathcal{O}(\epsilon^{2}).
\ee
With these approximations, it is now possible to write the approximate log-likelihood
\be
\begin{split}
\ln \mathcal{L}&=-\frac{1}{2}\left[ \Tr\ln \mathbf{P} +\mathbf{r}^{T}\mathbf{P}^{-1}\mathbf{r}-\mathbf{r}^{T}\mathbf{P}^{-1}\mathbf{S}_{c}\mathbf{P}^{-1}\mathbf{r}  \right]\\
&=-\frac{1}{2}\sum_{I}\left[ \Tr\ln P_{I}+r_{I}^{T}P_{I}^{-1}r_{I} \right]+\Omega\sum_{I,J}r_{I}^{T}P_{I}^{-1}S_{IJ}P_{J}^{-1}r_{J},
\end{split}
\ee
where again, we have used the notation that $\sum_{I,J}$ denotes a sum of all \emph{unique} pulsar pairs and $S_{IJ}$ is the \emph{amplitude-free} cross covariance matrix. The results of evaluating the first order likelihood for MDC's 1, 2, and 3 are shown in Figure \ref{fig:likeContours} and will be discussed in the next section.

\section{Executive Summary}
Here we will discuss our results in detail and use this information to make a decision as to what GWB signals are present in the data. As mentioned above, first we ran the noise estimation analysis on all pulsars of all datasets. For the purposes here we do not directly use the red noise information obtained from this search in subsequent analyses, however, we do record the white noise levels for all pulsars and find that the errorbars give a good estimate of the total white noise level. We will use these white noise values throughout our GWB analyses as the white noise components are not correlated with red noise components, thus, the measurements of the white noise are independent of red noise model assuming the data is well described by a two component noise model.

We have shown above that the first-order likelihood is designed for signals that are noise dominated. However, this is not the case for most residuals in the three datasets. For both MDC 2 and MDC 3 we have split the datasets up into thirds sorted by the white noise level. Using the open datasets and the noise estimation results as a gauge, we have found that the best 12 pulsars (sorted by the white noise level) is sufficiently noise dominated (across the 12 pulsar span) that the first-order likelihood performs well with no discernible biases. We have not used all 36 pulsars because we see biases in our results in the open datasets that we do not currently understand when using all 36 pulsars, and we have not chosen the worst 12 because they are noise dominated and show little evidence of a GWB. However, for MDC 1 all of the pulsars are in the large signal limit thus we simply choose a random group of 12 pulsars just to be consistent with the searches carried out for MDC's 2 and 3. We also use 12 pulsars in an attempt to keep the number of search parameters to a minimum (26 parameters in the case of 12 pulsars) and we have found that our analysis method using MultiNest is slowly convergent in large dimensional parameter spaces and in fact becomes computationally prohibitive in the 36 pulsar case. Other samplers are being explored but were not tested in time for this challenge.  

\begin{table*}[!h]
\caption{``best 12'' pulsars for each dataset.}
\centering
\begin{tabular}{lll}
\hspace{1cm}MDC1 & \hspace{1cm}MDC2 & \hspace{1cm}MDC3\\
\hline\hline
\hspace{1cm}J0030+0451 & \hspace{1cm}J1939+2134 & \hspace{1cm}J1939+2134\\ 
\hspace{1cm}J1738+0333 & \hspace{1cm}J0437-4715 & \hspace{1cm}J0437-4715\\ 
\hspace{1cm}J1741+1351 & \hspace{1cm}J1713+0747 & \hspace{1cm}J1713+0747\\ 
\hspace{1cm}J1744-1134 & \hspace{1cm}J1909-3744 & \hspace{1cm}J1909-3744\\ 
\hspace{1cm}J1751-2857 & \hspace{1cm}J1744-1134 & \hspace{1cm}J1744-1134\\ 
\hspace{1cm}J1853+1303 & \hspace{1cm}J1910+1256 & \hspace{1cm}J1910+1256\\ 
\hspace{1cm}J1857+0943 & \hspace{1cm}J1853+1303 & \hspace{1cm}J1853+1303\\ 
\hspace{1cm}J1732-5049 & \hspace{1cm}J1955+2908 & \hspace{1cm}J1955+2908\\ 
\hspace{1cm}J1909-3744 & \hspace{1cm}J1741+1351 & \hspace{1cm}J1741+1351\\ 
\hspace{1cm}J1918-0642 & \hspace{1cm}J1640+2224 & \hspace{1cm}J1640+2224\\ 
\hspace{1cm}J1939+2134 & \hspace{1cm}J1600-3053 & \hspace{1cm}J1600-3053\\ 
\hspace{1cm}J1955+2908 & \hspace{1cm}J1738+0333 & \hspace{1cm}J1738+0333\\ 
\hline
\label{tab:psrs}
\end{tabular}
\end{table*}

With these datasets now chosen we have run our first-order likelihood assuming three different models:
\begin{enumerate}

\item Stochastic GWB with amplitud$A$, spectral index $\gamma_{\rm gw}$ with Hellings-Downs correlation coefficients plus intrinsic red noise with a power law red noise with amplitude $A_{i}$ and spectral index $\gamma_{i}$ and white noise given by the noise estimation estimates. We use flat priors in the GW amplitude and spectral index parameters and the noise amplitude and spectral index with $A\in (0,5\times10^{-13})$, $\gamma_{\rm gw}\in[1,7]$, $A_{i}\in(0,5\times 10^{-13})$\footnote{Here we use a red noise amplitude that is analogous to the traditional GW strain amplitude}, and $\gamma_{i}\in [1,7]$. This results in a 2+$2M$ dimensional search, where $M$ is the number of pulsars used (two parameters for the GWB and 2$M$ for the intrinsic noise parameters).

\item Null hypothesis where all red noise processes are intrinsic. Again we assume an intrinsic amplitude $A_{i}$ and spectral index $\gamma_{i}$ with noise estimated white noise parameters and no GWB signal. This results in a $2M$ parameter search.

\item Common red noise signal with amplitude $\mathcal{B}_{\rm common}$ and spectral index $\gamma_{\rm common}$ but no spatial correlations and intrinsic red noise parameters as described above. Again, this results in  a $2+2M$ parameter search.
\end{enumerate}
For each model we evaluate the Bayesian evidence and then compute the Bayes factors $\mathcal{B}_{0}$ and $\mathcal{B}_{\rm common}$, which compare the GWB hypothesis to the null hypothesis and uncorrelated common red noise hypothesis, respectively. All of these searches were run on the Nemo Computing Cluster at UWM using the MultiNest \citep{fhb09} algorithm. Typical runtimes using 20 CPUs for each dataset (12 pulsars) are around 2 hours.

\begin{table*}
\caption{Table of results from many analysis methods. Here we have estimates of the GWB amplitude $A$ and the corresponding SNR using the zero-order optimal statistic and iterative optimal statistic methods described above. The uncertainty on $A$ represent the 1-$\sigma$ confidence levels.  We have also listed the results of our first-order likelihood method, also described above. We quote the maximum of the marginalized posterior distribution for both the GWB amplitude, $A$, and the spectral index $\gamma$ as well as the natural logarithm of the two Bayes factors described in the text. The uncertainties on $A$ and $\gamma$ represent the 68\% credible regions.  }
\centering
\begin{tabular}{p{0.09\textwidth}ccccccccc}
\\
& \multicolumn{2}{c} {Zero Order OS$^{a}$} \hspace{1cm} & \multicolumn{2}{c} {Iterative OS}\hspace{1cm} & & \multicolumn{4}{c} {First-Order Likelihood$^{b}$} \\
 \hline\hline
  & $A[\times 10^{-14}]$ & SNR\hspace{1cm} & $A[\times 10^{-14}]$ & SNR\hspace{1cm} & & $A[\times 10^{-14}]$ & $\gamma$ & $\ln\,\mathcal{B}_{0}$ & $\ln\,\mathcal{B}_{\rm common}$\\
  \hline
MDC 1 & $0.45$ & 4.3\hspace{1cm} & $0.92\pm0.2$ & 9.2\hspace{1cm} & &$1.2^{+0.19}_{-0.16}$ & $4.26^{+0.22}_{-0.27}$ & 33.5 & 16.8 \\
MDC 2 & $1.8$ & 2.0\hspace{1cm} & $4.3\pm1.4$ & 9.5\hspace{1cm} & & $6.87^{+0.40}_{-0.65}$ & $4.16^{+0.18}_{-0.15}$ & 153.4 & 10.5 \\
MDC 3 & $0.15$ & 1.2\hspace{1cm} & $0.63\pm0.38$ & 5.5\hspace{1cm} & & $0.88^{+0.20}_{-0.22}$ & $3.58^{+0.48}_{-0.44}$ & 31.6 & 20.6 \\
\hline
\multicolumn{9}{l}{ \footnotesize{$^{a}$ An estimate of the amplitude is required to fully calculate the uncertainty, thus we do not include it here.}}\\
\multicolumn{9}{l}{ \footnotesize{$^{b}$ Note that these results are only for the ``best 12'' pulsars.}}\\
\label{tab:results}
\end{tabular}
\end{table*}

\begin{figure*}[!h]
  \begin{center}
	\includegraphics[scale=0.8]{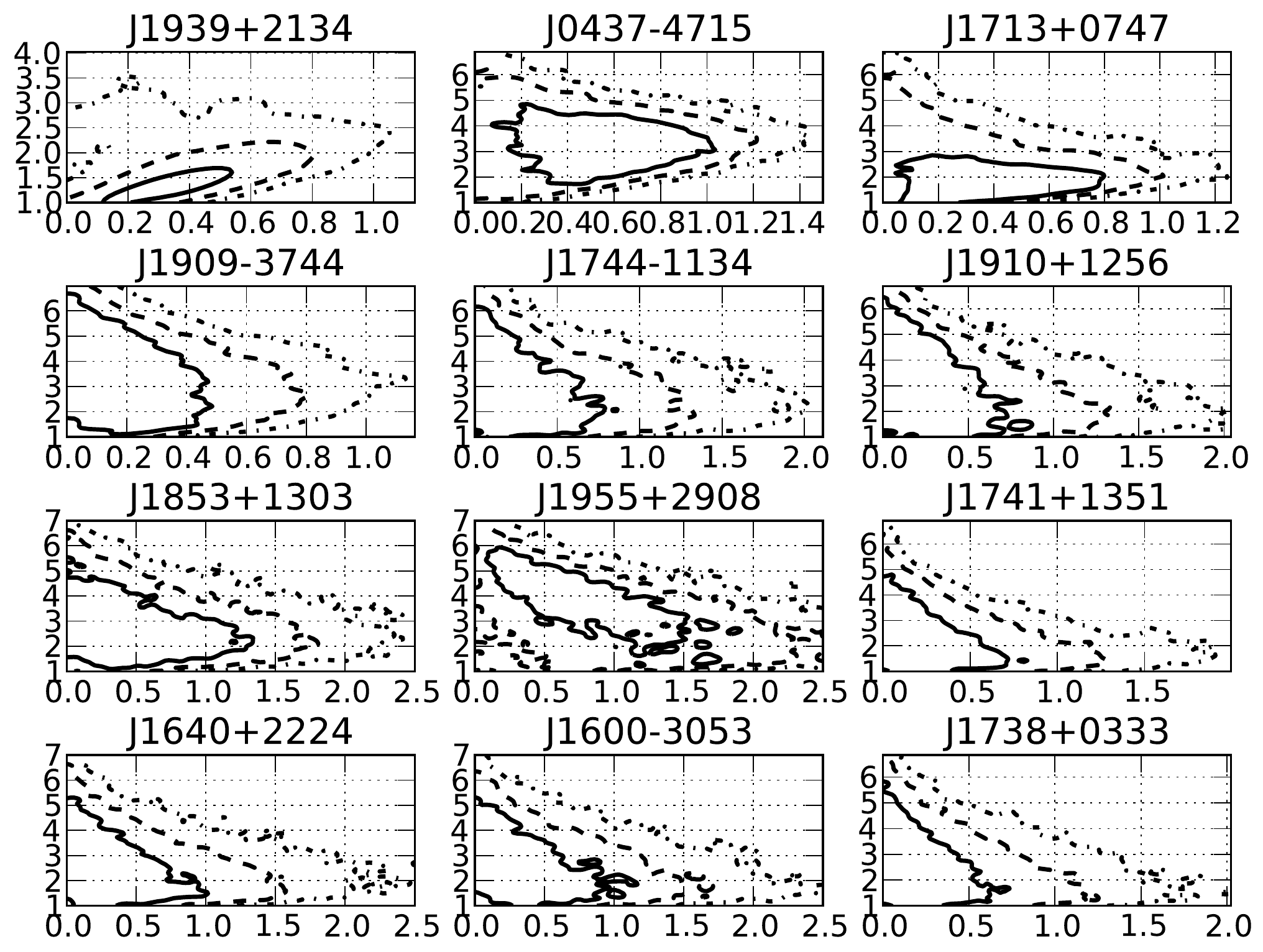}
  \end{center}
  \caption{Noise contours for the ``best 12'' pulsars of MDC3. The $x$-axis is the amplitude measured in GW strain units normalized by $10^{-14}$ and the $y$-axis is the power spectral index. The solid, dashed, and dot-dashed lines represent the 1, 2, and 3 sigma confidence contours, respectively.}
\label{fig:red}
\end{figure*}

The results of the first order likelihood run on the ``best 12'' pulsars is shown in Table \ref{tab:results} and the list of pulsars used in each dataset is shown in table \ref{tab:psrs}. Here we quote the maximum of the marginalized posterior distribution for both the amplitude (written in characteristic strain units) and spectral index as well as the 68\% credible regions and the logarithm of the two Bayes factors, $\mathcal{B}_{0}$ and $\mathcal{B}_{\rm common}$.  Firstly, we note that all three datasets (using the ``best 12'' pulsars) have large Bayes factors in favor of a GWB model over a model with just \emph{intrinsic} red noise\footnote{Remember, a Bayes factor $>10$ is strong evidence against the competing hypothesis.}. We also note that all 3 datasets also strongly favor a GWB model over an uncorrelated common red noise signal. As we will note shortly, we can use the optimal statistic in combination with the first-order likelihood to make a model decision in situations where the evidence for a particular model is weak. We also see that the most likely spectral index is consistent with $13/3$ (SMBHB spectrum) for all three datasets. We also note that there was no evidence for red noise in MDC1 or MDC2, however, we did find evidence for intrinsic red noise in closed dataset 3. Although there is evidence for red noise, we cannot accurately characterize it in many cases. For PSRs J1939+2134, J0437-4715 and J1955+2908, the 1-sigma contours do are not consistent with red noise of 0 amplitude, however for all others we are unable to characterize the red noise. Contour plots of the red noise parameters for MDC3 are shown in Figure \ref{fig:red}.

We have also run both the zero-order and iterative optimal statistics on the \emph{full} datasets. Typical run times are on the order of 5 minutes on a modern workstation. As mentioned above, for the zero order OS we use the noise estimation results shown in Figures \ref{fig:noiseMDC1}--\ref{fig:noiseMDC3} to construct the auto-covariance matrices. The amplitude (written in characteristic strain units) and corresponding SNR are shown in the second and third columns of Table \ref{tab:results}. We see that there is a significant detection for MDC 1 and marginal detections for MDC's 2 and 3 using this method. Intuitively, this makes sense because there is more white noise in pulsars from MDC's 2 and 3 and thus a larger spread in the estimated amplitudes and spectral indices of the GWB (based on the first order likelihood results we assume that any measured red noise is due to the GWB) and thus a more biased result from the optimal statistic. When we now use the iterative method we make confident detections for all three datasets and recover amplitudes that are consistent with the first-order likelihood at the 95\% level or better.  



\end{document}